\documentclass{sig-alternate}
\usepackage{color}
\usepackage{multirow}
\usepackage{array}

\newcolumntype{L}[1]{>{\raggedright\let\newline\\\arraybackslash\hspace{0pt}}m{#1}}
\newcolumntype{C}[1]{>{\centering\let\newline\\\arraybackslash\hspace{0pt}}m{#1}}
\newcolumntype{R}[1]{>{\raggedleft\let\newline\\\arraybackslash\hspace{0pt}}m{#1}}

\def\sharedaffiliation{%
\end{tabular}
\begin{tabular}{c}}

\numberofauthors{3} 
\author{
\alignauthor
Xiao Yang\\
\alignauthor
Craig Macdonald\\
\alignauthor 
Iadh Ounis\\
\sharedaffiliation
	\email{\{firstname.lastname\}@glasgow.ac.uk}\\
       \affaddr{University of Glasgow, UK}
}

\begin{document}



\title{Using Word Embeddings in Twitter Election Classification}

\date{1 May 2016}

\maketitle

\begin{abstract}
Word embeddings and convolutional neural networks (CNN) have attracted extensive attention in various classification tasks for Twitter, e.g.\ sentiment classification. However, the effect of the configuration used to train and generate the word embeddings on the classification performance has not been studied in the existing literature. In this paper, using a Twitter election classification task that aims to detect election-related tweets, we investigate the impact of the background dataset used to train the embedding models, the context window size and the dimensionality of word embeddings on the classification performance. By comparing the classification results of two word embedding models, which are trained using different background corpora (e.g.\ Wikipedia articles and Twitter microposts), we show that the background data type should align with the Twitter classification dataset to achieve a better performance. Moreover, by evaluating the results of word embeddings models trained using various context window sizes and dimensionalities, we found that large context window and dimension sizes are preferable to improve the performance. Our experimental results also show that using word embeddings and CNN leads to statistically significant improvements over various baselines such as random, SVM with TF-IDF and SVM with word embeddings.
\end{abstract}

\section{Introduction}
Word embeddings have been proposed to produce more effective word representations. For example, in the \textit{Word2Vec} model \cite{Mikolov2013a}, by maximising the probability of seeing a word within a fixed context window, it is possible to learn for each word in the vocabulary a dense real valued vector from a shallow neural network. As a consequence, similar words are close to each other in the embedding space \cite{Bengio2003, Collobert2011, Mikolov2013a}. The use of word embeddings together with convolutional neural networks (CNN) has been shown to be effective for various classification tasks such as sentiment classification on Twitter \cite{Ebert2015, Severyn2015}. However, the effect of the configuration used to generate the word embeddings on the classification performance has not been studied in the  literature. Indeed, while different background corpora (e.g.\ Wikipedia, GoogleNews and Twitter) and parameters (e.g.\ context window and dimensionality) could lead to different word embeddings, there has been little exploration of how such background corpora and parameters affect the classification performance. 

In this paper, using a dataset of tweets collected during the Venezuela parliamentary election in 2015, we investigate the use of word embeddings with CNN in a new classification task, which aims to identify those tweets that are related to the election. Such a classification task is challenging because election-related tweets are usually ambiguous and it is often difficult for human assessors to reach an agreement on their relevance to the election \cite{Bermingham2011}. For example, such tweets may refer to the election implicitly without mentioning any political party or politician. In order to tackle these challenges, we propose to use word embeddings to build richer vector representations of tweets for training the CNN classifier on our election dataset. 

We thoroughly investigate the effect of the background corpus, the context window and the dimensionality of word embeddings on our election classification task. Our results show that when the type of background corpus aligns with the classification dataset, the CNN classifier achieves statistically significant improvements over the most effective classification baseline of SVM with TF-IDF on our task. We also show that word embeddings trained using a large context window size and dimension size can help CNN to achieve a better classification performance. Thus, our results suggest indeed that the background corpus and parameters of word embeddings have an impact on the classification performance. Moreover, our results contradict the findings of different tasks such as dependency parsing \cite{Bansal2014} and named entity recognition (NER) \cite{Godin2015} where a smaller context window is suggested. Such a contradiction suggests that the best setup of parameters such as the context window and dimensionality might differ from a task to another.

In the remainder of this paper, we briefly explain the related work in Section 2. We describe and illustrate the CNN architecture used for our classification task in Section 3. In Section 4, we describe our dataset and the experimental setup. In Section 5, we discuss the impact of two background corpora (Wikipedia articles and Twitter microposts) on the effectiveness of the learned classifier. In Section 6, we investigate the impact of the context window size and dimensionality of word embeddings on the classification performance. We provide concluding remarks in Section 7.

\section{Related work}
\label{sec:related work}
\begin{figure*}[ht]
\centering
\includegraphics[width=0.5\linewidth]{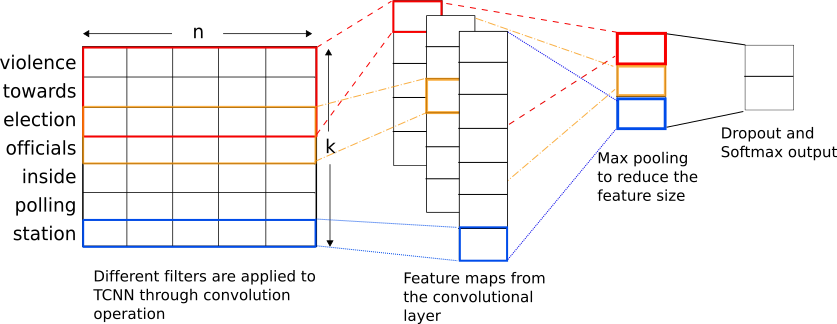}
\caption{Convolutional neural network architecture for tweet classification. Adapted from \cite{Kim2014}.}
\vspace{-2mm}
\label{fig:cnn graph}
\end{figure*}
A number of studies have already shown that the context window and dimensionality of the used word embedding vectors could affect performance in tasks such as dependency parsing \cite{Bansal2014} and named entity tagging \cite{Godin2015}. For instance, using publicly available corpora such as Wall Street Journals and Wikipedia, Bansal et al.\ \cite{Bansal2014} investigated \textit{Word2Vec} word embeddings in the dependency parsing task, which aims to provide a representation of grammatical relations between words in a sentence. By only varying the context window size from 1 to 10, their results on the accuracy of part-of-speech (POS) tagging showed that the context window size of \textit{Word2Vec} could affect the type of the generated word embedding. In particular, they observed that a smaller context window gives a better performance on accuracy. In the named entity recognition (NER) task, Godin et al.\ \cite{Godin2015} investigated three context window sizes $w$ of $w=\{1, 3, 5\}$ based on the accuracy of NER tagging. Their results also reached the same conclusion, namely that a smaller context window gives a better performance using the \textit{Word2Vec} word embeddings when the model is trained from a large Twitter corpus containing 400 million tweets. 

Using a subset of the semantic-syntactic word relationship test set, Mikolov et al.\ \cite{Mikolov2013a} investigated the dimensionality of the \textit{Word2Vec} word embeddings and the size of background data. In the test set, word pairs are grouped by the type of relationship. For example ``brother-sister'' and ``grandson-granddaughter'' are in the same relationship of ``man-woman''. The accuracy is measured such that given a word pair, another word pair with the correct relationship should be retrieved. Using this accuracy measure, they noted that at some point increasing the dimensionality or the size of background data only provides minor improvements. Thus, they concluded the dimensionality and background data size should be increased together \cite{Mikolov2013a}. However, Mikolov et al.\ \cite{Mikolov2013a} only investigated the \textit{Word2Vec} parameters using the GoogleNews background corpus. 

The aforementioned studies provide a useful guide about the effect of the word embeddings configuration on performance in the specific applications they tackled, but their findings were obtained on tasks different from Twitter classification tasks. Hence, the question arises as whether such findings will generalise to classification tasks on Twitter, which is the object of our study in this paper. 

In fact, there is little work in the literature tackling the task of election classification on Twitter. However, similar classification tasks such as Twitter sentiment classification have been well studied \cite{Ebert2015,Severyn2015,Tang2014a}. In particular, word embeddings were recently used to build effective tweet-level representations for Twitter sentiment classification \cite{Severyn2015,Tang2014a}. For instance, in the Semeval-2015 Twitter Sentiment Analysis challenge, Severyn et al.\ \cite{Severyn2015} proposed to use word embeddings learned from two Twitter corpora to build the vector representations of tweets. Using the \textit{Word2Vec} model, default parameter values such as context window size 5 and dimensionality 100 were applied to train the word embedding. In their approach, one Twitter background corpus (50 million tweets) was used to train the word embedding, while another one (10 million tweets) containing positive and negative emoticons was used to refine the learned word embeddings using the proposed CNN classifier. The CNN classifier was then trained on the Semeval-2015 Twitter sentiment analysis dataset, which contains two subsets: phrase-level dataset and message-level dataset. Each subset contains 5K+ and 9K+ training samples, respectively. The official ranking in Semeval-2015 showed that this system ranked 1st and 2nd on the phase-level dataset and the message-level dataset, respectively. However, Severyn et al.\ \cite{Severyn2015} focused on refining the word embeddings by using another Twitter corpus with emoticons to learn sentiment information, but did not study the impact of the background corpus and the chosen parameters on the classification performance. 

In another approach based on the word embeddings model proposed by Collobert et al.\ \cite{Collobert2008}, Tang et al.\ \cite{Tang2014a} proposed a variation to learn sentiment-specific word embeddings (SSWE) from a large Twitter corpus containing positive and negative emoticons. Tang et al.\ \cite{Tang2014a} empirically set the context window size to 3 and the embedding dimensionality to 50. The Semeval-2013 Twitter sentiment analysis dataset, which contains 7K+ tweets was used to evaluate the effectiveness of their proposed approach. Compared to the top system of the Semeval-2013 Twitter Sentiment Analysis challenge, their approach of using an SVM classifier with SSWE outperformed the top system on the F1 measure. However, only the Twitter background corpus was used by Tang et al.\ \cite{Tang2014a}, which contains 10 million tweets with positive and negative emoticons. On the other hand, the parameters of word embeddings such as the context window and dimensionality were not studied by Tang et al.~\cite{Tang2014a}, nor in the existing literature for Twitter classification tasks. As such, in this paper, we conduct a thorough investigation of word embeddings together with CNN on a Twitter classification task and explore the impact of both the background corpus, the context window and the dimensionality of word embeddings on the classification performance.

\section{The CNN model}
\label{sec: cnn}
For our Twitter election classification task, we use a simple CNN architecture described by Kim \cite{Kim2014} as well as the one proposed by Severyn et al. \cite{severyn2015d} and highlighted in Fig.\ \ref{fig:cnn graph}. It consists of a convolutional layer, a max pooling layer, a dropout layer and a fully connected output layer. Each of these layers is explained in turn. 

\textbf{Tweet-level representation.}
The inputs of the CNN classifier are preprocessed tweets that consist of a sequence of words. Using word embeddings, tweets are converted into vector representations in the following way. Assuming $w_i \in \mathbb{R}^n$ to be the $n$-dimensional word embeddings vector of the $i$th word in a tweet, a tweet-level representation is obtained by looking up the word embeddings and concatenating the corresponding word embeddings vectors of the total $k$ words:
\begin{equation}
TCNN = w_1 \oplus w_2 \oplus \cdots \oplus w_k
\label{eq:cnn concatenating}
\end{equation}
where $\oplus$ denotes the concatenation operation \cite{Kim2014}. For training purposes, short tweets in our dataset are padded to the length of the longest tweet using a special token. Hence the total dimension of the vector representation $TCNN$ is always $k \times n$. Afterwards, the tweet-level representation will feed to the convolutional layer.

\textbf{Convolutional layer.} The convolution operation helps the network to learn the important words no matter where they appear in a tweet \cite{Severyn2015}. In this layer, the filter $F_i \in \mathbb{R}^{m \times n}$ with different sizes of $m$ are applied to the tweet-level representation $TCNN$. By varying the stride $s$ \cite{krizhevsky2012}, we can shift the filters across $s$ word embeddings vectors at each step. By sliding the filters over $m$ word vectors in $TCNN$ using stride $s$, the convolution operation produces a new feature map $c_i$ for all the possible words in a tweet:
\begin{equation}
c_i = f(F_i \cdot TCNN_{i:i+m-1} + b_i)
\label{eq:cnn conv}
\end{equation}
where $i:i+m-1$ denotes the word vectors of word $i$ to word $i+m-1$ in $TCNN$. $b_i$ is the corresponding bias term that is initialised to zero and learned for each filter $F_i$ during training. In Eq.\ \eqref{eq:cnn conv}, $f$ is the activation function. In this CNN architecture, we used a rectified linear function (ReLU) as $f$. No matter whether the input $x$ is positive or negative, the ReLU unit ensures its output (i.e.\ $c_i$) is always positive as defined by $f = max(0,x)$. 

\textbf{Max pooling layer.}
All the feature maps $c_i$ from the convolutional layer are then applied to the max pooling layer where the maximum value $c_i^{max}$ is extracted from the corresponding feature map. Afterwards, the maximum values of all the feature maps $c_i$ are concatenated as the feature vector of a tweet. 

\textbf{Dropout layer.}
Dropout is a regularization technique that only keeps a neuron active with some probability $p$ during training \cite{Kim2014}. After training, $p=1$ is used to keep all the neurons active for predicting unseen tweets. Together with the $L2$ regularization, it constraints the learning process of the neural networks by reducing the number of active neurons. 

\textbf{Softmax Layer.}
The outputs from the dropout layer are fed into the fully connected softmax layer, which transforms the output scores into normalised class probabilities \cite{Kim2014}. Using a cross-entropy cost function, the ground truth labels from human assessors are used to train the CNN classifier for our Twitter election classification task.

During training, the weights of each layer are updated according to the loss between the prediction and the target. Once a CNN classifier is trained from a training set, all of its parameters and learned weights are saved into binary files that can be loaded to classify unseen tweets using the same procedures explained in this section.

\section{Experimental setup}
\label{sec:experimental setup}
In this paper, we argue that the types of background corpora as well as the parameters of \textit{Word2Vec} model could lead to different word embeddings and could affect the performance on Twitter classification tasks. In the following sections, experiments are tailored to conduct a thorough investigation of word embeddings together with CNN on a Twitter classification task and to explore the impact of the background corpora (Section~\ref{sec5}), the context window and the dimensionality of word embeddings (Section~\ref{sec6}) on the classification performance. The remainder of this section details our dataset (Section \ref{sec:dataset}), our experimental setup and used word embedding models (Section \ref{sec:word embedding}), baselines (Section \ref{sec:baselines}) and measures (Section \ref{sec:measures}). 

\subsection{Dataset}
\label{sec:dataset}
Our manually labelled election dataset is sampled from tweets collected about the 2015 Venezuela parliamentary election using the well-known pooling method \cite{sanderson2010}. It covers the period of one month before and after the election date (06/12/2015) in Venezuela. We use the Terrier information retrieval (IR) platform~\cite{Macdonald2012} and the DFReeKLIM~\cite{Amati2011} weighting model designed for microblog search to retrieve tweets related to 21 query terms (e.g.\ ``violencia'', ``eleccion'' and ``votar''). Only the top 7 retrieved tweets are selected per query term per day, making the size of the collection realistic for human assessors to examine and label the tweets. Sampled tweets are merged into one pool and judged by 5 experts who label a tweet as: ``Election-related'' or ``Not Election-related''. To determine the judging reliability, an agreement study was conducted using 482 random tweets that were judged by all 5 assessors. Using Cohen's $kappa$, we found a moderate agreement of 52\% between all assessors. For tweets without a majority agreement, an additional expert of Venezuela politics was used to further clarify their categories. In total, our election dataset consists of 5,747 Spanish tweets, which contains 9,904 unique words after preprocessing (stop-word removal \& Spanish Snowball stemmer). Overall, our labelled election dataset covers significant events (e.g.\ Killing of opposition politician Luis Diaz~\cite{bbc2015}) in the 2015 Venezuela parliamentary election. From the general statistics shown in Table~\ref{tb: stats}, we observe that the dataset is unbalanced; the majority class (Non-Election) has 1,000 more tweets than the minority class (Election).

\begin{table}[t]
\centering
\begin{tabular}{|c|c|c|c|c|}
\cline{2-5}
\multicolumn{1}{c|}{ }& Election& Non-Election & Total & \# Words\\
\hline
 Dataset &  2,274 &  3,474 &  5,747 & 9,904\\
 \hline
\end{tabular}
\caption{Statistics of the dataset used in the experiments. Negative class is the majority in the dataset.}
\vspace{-4mm}
\label{tb: stats}
\end{table}

\subsection{Word embeddings}\label{sec:word embedding}
The word embeddings used in this paper are trained from two different background corpora: a Spanish Wikipedia dump dated $02/10/2015$ (denoted \texttt{es-Wiki}) and a Spanish Twitter data (denoted \texttt{es-Twitter}) collected from the period of $05/01/2015$ to $30/06/2015$. Over 1 million Spanish articles are observed in \texttt{es-Wiki}. In \texttt{es-Twitter}, over 20 million Spanish tweets are collected by removing tweets with less than 10 words, hence the short and less informative tweets are not considered. For consistency, we apply the same preprocessing namely stop-word removal and stemmer (see Section \ref{sec:dataset}) to both of the background corpora. After the preprocessing, \texttt{es-Wiki} contains 436K unique words while \texttt{es-Twitter} has 629K unique words. Salient statistics are provided in Table \ref{tab: word coverage}. Indeed, by comparing the unique words in our election dataset with the words in \texttt{es-Wiki} and \texttt{es-Twitter}, we observe that 5,111 words in our dataset appear in \texttt{es-Wiki} while 6,612 words appear in \texttt{es-Twitter}. This shows that \texttt{es-Twitter} has a better word coverage on our election dataset. 

We use the \textit{Word2Vec} implementation in \textit{deeplearning4j} to generate a set of word embeddings by varying the context window size $W$ and the dimensionality $D$. We use the same context window sizes $W= \{1, 3, 5\}$ that were used by Godin et al.\ \cite{Godin2015}. For each context window $W$, we use three different dimension sizes $D =\{200, 500, 800\}$ to cover both of the low and high dimensionalities of the word embedding vectors, which were used by Mikolov et al.~\cite{Mikolov2013a}. Therefore, 9 word embeddings in total are generated by varying $W$ and $D$. For other parameters, we use the same values that were set by Mikolov et al.~\cite{Mikolov2013a}: We set the batch size to $50$, negative sampling to $10$, minimum word frequency to $5$ and iterations to $5$. As suggested by Kim \cite{Kim2014}, for a word not appearing in a word embeddings (also known as {\em out-of-vocabulary OOV}), we generate its vector by sampling each dimension from the uniform distributions $U_i[m_i - s_i, m_i + s_i]$, where $m_i$ and $s_i$ are the mean and standard deviation of the $i$th dimension of the word embeddings.

\begin{table}
\centering
\begin{tabular}{|c|c|c|}
\cline{2-3}
\multicolumn{1}{c|}{ }& \texttt{es-Wiki} & \texttt{es-Twitter} \\
\hline
\# Documents & 1M+ & 20M+\\
\hline
 Vocabulary Size & 436K & 629K \\
 \hline
 Word Coverage Count &  5,111 & 6,612\\
\hline
 Word Coverage Rate &  51\% &  66\%  \\
\hline
\end{tabular}
\caption{Statistics of the background corpora and words coverage on the election dataset.}
\vspace{-5mm}
\label{tab: word coverage}
\end{table}

\subsection{Baselines}
\label{sec:baselines}
To evaluate the CNN classifiers and word embeddings, we use three baselines, namely:

\textbf{\textit{Random classifier}}: The random classifier simply makes random predictions to the test instances.

\textbf{\textit{SVM with TF-IDF}} (SVM+TFIDF): As a traditional weighting scheme, TF-IDF is used in conjunction with an SVM classifier for the Twitter election classification.

\textbf{\textit{SVM with word embeddings}} (SVM+WE): We use a similar scheme that was used by Wang et al.~\cite{Wang2015} to build the tweet-level representation for the SVM classifiers. The vector representation (i.e.\ $TWE$) of a tweet is constructed by averaging the word embedding vectors along each dimension for all the words in the tweet:
\begin{equation}
TWE = \sum_{i=1}^{k} w_i/k
\label{eq: sum we}
\end{equation}
where $k$ is the number of words in a tweet and $w_i \in \mathbb{R}^n$ denotes the word embedding vector of the $i$th word. The vector representation of each tweet has exactly the same dimension as the word embedding vector $w_i$, which is the input of an SVM classifier.

\subsection{Hyperparameters and measures}
\label{sec:measures}
For all the experiments, we use 3 filter sizes $m = \{1, 2, 3\}$, stride $s=1$ and dropout probability $p=0.5$ for our CNN classifier, following the settings used by Kim~\cite{Kim2014}. For each filter size, $200$ filters are applied to the convolutional layer and therefore $600$ feature maps are produced in total. For the SVM classifier, we use the default parameter $c=1$ for the \textit{LinearSVC} implementation in \textit{scikit-learn}\footnote{An open source machine learning library in Python.} \cite{Pedregosa2011}.

To train the classifiers and evaluate their performances on our dataset, we use a 5-fold cross validation, such that in each fold, 3 partitions are used for training, 1 partition for validation and 1 partition for test. We stop the training process when the classification accuracy on the validation partition declines. Afterwards, the overall performance on the test instances is assessed by averaging the scores across all folds. We report effectiveness in terms of classification measures, precision (denoted $P$), recall (denoted $R$) and F1 score (denoted $F1$).

\section{Effect of the background \\corpora}\label{sec5}

\begin{table}
\centering
\begin{tabular}{|c|c|c|c|c|c|c|}
\hline
\multirow{2}{*}{\textbf{Classifier}} & \multicolumn{3}{c}{\texttt{es-Wiki}} & \multicolumn{3}{|c|}{\texttt{es-Twitter}} \\
\cline{2-7}
& P & R & F1 &P & R & F1\\
\hline
SVM+WE & \textbf{74.9} & 68.0 & 71.3 & 74.3 & \textbf{70.1} & \textbf{72.1} \\
\hline
CNN  & \textbf{81.6} & 70.7 &  75.8 & 80.9 & \textbf{72.2} & \textbf{76.3}\\
\hline
\end{tabular}
\caption{Classification results by using the background corpora \texttt{es-Wiki} and \texttt{es-Twitter}.}
\vspace{-2mm}
\label{tab: data type}
\end{table}

Due to the noisy nature of Twitter data, Twitter posts can often be poor in grammar and spelling. Meanwhile, Twitter provides more special information such as Twitter handles, HTTP links and hashtags which would not appear in common text corpora. In order to infer whether the type of background corpus could benefit the Twitter classification performance, we compare the two background corpora of \texttt{es-Wiki} and \texttt{es-Twitter}. By considering the various experimental results in \cite{Bansal2014,Godin2015,Mikolov2013a}, the context window size of $5$ is said to give a good performance. Thus, in this experiment we set the context window to $5$ and the dimensionality to $500$ for both word embeddings. 

The classification results are shown in Table~\ref{tab: data type} where the first column shows the classifiers we used. In other columns, we report three measures for both the background corpora \texttt{es-Wiki} and \texttt{es-Twitter}. Since the SVM+TFIDF and random classifier do not use the background corpus, they are not listed in Table \ref{tab: data type}. For each classifier, the best scores are highlighted in bold. From Table~\ref{tab: data type}, we observe that when the type of background corpus aligns with our Twitter election dataset, the performance is better for both the SVM+WE and CNN classifiers on Recall and F1 score. In particular, the improvement on recall suggests that \texttt{es-Twitter} represents the characteristics of Twitter posts better than the \texttt{es-Wiki} corpus. 

As shown in the statistics of the two background corpora (Table~\ref{tab: word coverage}), 66\% of the vocabulary of our election dataset appears in \texttt{es-Twitter} while only 51\% appears in \texttt{es-Wiki}. By removing the words covered by both background corpora, we observe that 1,527 unique words are covered by \texttt{es-Twitter} but not covered by \texttt{es-Wiki}. However, there are only 26 unique words that are covered by \texttt{es-Wiki} only. Table~\ref{tab: coverage stat} categorises the words only found in \texttt{es-Twitter}, which are mostly words unique to Twitter, such as Twitter handles and hashtags. This explains why \texttt{es-Twitter} works better with our Twitter election dataset. The other 374 words are mainly incorrect spellings and elongated words such as ``bravoooo'', ``yaaaa'' and ``urgenteeeee'', which occur more often in Twitter than in other curated types of data such as Wikipedia and News feeds. Our finding on the vocabulary coverage further validates our results shown in Table \ref{tab: data type}. Thus, the results may generalise to similar Twitter classification tasks that also deal with Twitter posts. In summary, we find that aligning the type of background corpus with the classification dataset leads to better feature representations, and hence a more effective classification using the CNN classifier. 

\begin{table}
\centering
\begin{tabular}{|c|c|c|c|}
\hline
\textbf{Twitter handles} & \textbf{Hashtags} & \textbf{Others} & \textbf{Total} \\
\hline
 818 &  335 &  374 &  1,527\\
 \hline
\end{tabular}
\caption{Statistics of the vocabulary only covered by \texttt{es-Twitter}.}
\vspace{-4mm}
\label{tab: coverage stat}
\end{table}

\section{Effect of word embeddings \\parameters}\label{sec6}
In this section, we attempt to investigate the effect of parameters (e.g.\ context window and dimensionality) for the Twitter election classification task. Since \texttt{es-Twitter} gives a better performance, we only use word embeddings generated from \texttt{es-Twitter} only. Table~\ref{tab: precision results}(a) shows the results of our three baselines, while Table~\ref{tab: precision results}(b) shows the results of classifiers using word embeddings, namely SVM with word embeddings (SVM+WE) and CNN. In Table~\ref{tab: precision results}(b), the measurements for SVM+WE and CNN are arranged by the dimensionality and context window size of word embeddings. For each row of $W1$, $W3$ and $W5$, Table \ref{tab: precision results}(b) shows results for context window sizes of $W=\{1,3,5\}$ along each dimension sizes of $D=\{200,500,800\}$. The best overall scores are highlighted in bold.

We first compare the results of the CNN classifiers to the random baseline and the SVM+WE baseline. Clearly, the CNN classifiers outperform these two baselines across all measures. By comparing CNN classifiers to the best baseline SVM+TFIDF, the CNN classifiers consistently outperform the SVM+TFIDF baseline on precision and F1 score. In particular, when $W=5$ and $D=800$, the CNN classifier achieves the best scores on all the metrics, which shows the effectiveness of convolution neural networks with word embeddings in the Twitter election classification task. In order to validate whether the best CNN classifiers significantly outperforms the best baseline SVM+TFIDF, the non-parametric McNemar's test is used to conduct a statistical test as suggested by Dietterich \cite{dietterich1998} for a reliable and computational inexpensive comparison. Our statistical test result shows that the two-tailed $p$-value is 0.0042, which means the difference between CNN and SVM+TFIDF is considered to be statistically significant.

In Table \ref{tab: precision results}(b), where both approaches use word embeddings, we observe that SVM+WE and CNN show different preferences in word embeddings dimensionality. When using SVM+WE, a smaller dimension size and larger context window size (for example $W5$ and $D200$) give a better performance on F1 score and precision. However, the CNN classifier prefers both large context window size and dimension size. Therefore, when using a large context window size, word embeddings with higher dimensionality are likely to have a better performance (for example $W5$ and $D800$). The simple scheme used in SVM+WE is problematic with high dimensional word embeddings. Simply combining all the word vectors of a Twitter post may excessive ambiguity about the topic of the post, particularly as not all the words are meaningful for classification. Hence, this scheme may hurt the semantic representation~\cite{Wang2015}. As the dimensionality increases, this could introduce further ambiguities and lead to degraded performance in our Twitter election classification task. Nevertheless, results of both SVM+WE and CNN suggest that a higher context window size is most appropriate for our task. 

Compared to the studies on other tasks such as named entity recognition (NER) and dependency parsing (see Section \ref{sec:related work}), our results differ from their conclusions that ``a smaller context window size gives a better performance'' \cite{Bansal2014,Godin2015}. Such a contradiction suggests that the best setup of parameters such as context window and dimensionality might differ from a task to another. In summary, for the Twitter election classification task using CNNs, word embeddings with a large context window and dimension size can achieve statistically significant improvements over the most effective classification baseline of SVM with TF-IDF.

\section{Conclusion}
\begin{table*}[tb]
\centering
\setlength{\tabcolsep}{3pt}
\setlength\extrarowheight{3pt}
\begin{tabular}{ c| ccc | ccc| ccc | ccc |ccc | ccc |}
\multicolumn{19}{c}{(a) Results of random classifier, SVM with TF-IDF (SVM+TFIDF) and SVM with word embeddings (SVM+WE)}\\
\cline{8-16}
\multicolumn{7}{c}{} & \multicolumn{3}{|c}{Precision} &\multicolumn{3}{|c}{Recall} & \multicolumn{3}{|c|}{F1 score} & \multicolumn{3}{c}{}\\
\cline{5-16}
\multicolumn{4}{c}{}&\multicolumn{3}{|c}{Random} & \multicolumn{3}{|c}{38.6} &\multicolumn{3}{|c|}{28.5} & \multicolumn{3}{c|}{38.5} & \multicolumn{3}{c}{}\\
\cline{5-16}
\multicolumn{4}{c}{}&\multicolumn{3}{|c}{SVM+TFIDF} & \multicolumn{3}{|c}{76.0} &\multicolumn{3}{|c|}{\textbf{73.7}} & \multicolumn{3}{c|}{\textbf{74.8}} & \multicolumn{3}{c}{}\\
\cline{5-16}
\multicolumn{4}{c}{}&\multicolumn{3}{|c}{SVM+WE$^*$} & \multicolumn{3}{|c}{\textbf{77.4}} &\multicolumn{3}{|c|}{68.4} & \multicolumn{3}{c|}{72.6} & \multicolumn{3}{c}{}\\
\cline{5-16}
\multicolumn{19}{c}{(b) Results of SVM with word embedding (SVM+WE) and CNN}\\
\cline{2-19}
& \multicolumn{6}{c}{\textbf{D200}} & \multicolumn{6}{|c}{\textbf{D500}} & \multicolumn{6}{|c|}{\textbf{D800}}\\
\cline{2-19}
& \multicolumn{3}{c}{SVM+WE} & \multicolumn{3}{c|}{CNN} &\multicolumn{3}{c}{SVM+WE} & \multicolumn{3}{c|}{CNN} & \multicolumn{3}{c}{SVM+WE} &\multicolumn{3}{c|}{CNN} \\
\cline{2-19}
& P & R & F1 & P & R & F1 & P & R & F1 & P & R & F1 & P & R & F1 & P & R & F1\\
\hline
\multicolumn{1}{|c|}{{\rotatebox{90}{\textbf{W1}}}}& 75.9 & 66.8  & 71.0 & 81.2 &  71.6 & 76.1 & 74.2 & 68.8 & 71.4 & 80.7 & 71.9 & 76.0 & 71.7 & 69.8 & 70.7 & 80.8  & 71.6 & 75.9\\
\hline
\multicolumn{1}{|c|}{{\rotatebox{90}{\textbf{W3}}}}& 77.5 & 67.7 & 72.2 & 81.2 &  72.0 & 76.3 & 75.4 & 68.9 & 72.0 &80.9 & 71.6 & 75.9 & 71.8 & 68.9 &70.3 &  81.3  & 72.2 & 76.5\\
\hline
\multicolumn{1}{|c|}{{\rotatebox{90}{\textbf{W5}}}} & 77.4 &  68.4 & 72.6  & $82.0^\dag$ &  $71.5^\dag$ & $76.4^\dag$  & 74.3 & 70.1 & 72.1 &  $80.9^\dag$ & $72.2^\dag$ & $76.3^\dag$ & 71.5 & 69.3 & 70.4  & $\textbf{80.6}^\dag$  & $\textbf{74.0}^\dag$ & $\textbf{77.1}^\dag$\\
\hline

\end{tabular}
\caption{Results of our baselines and CNN models in Twitter election classification task. $W1$ means context window size $1$ and $D200$ denotes word embeddings dimension size $200$. $^*$ denotes the setting of SVM+WE that exhibits the highest F1 score (W5 and D200) from (b). $^\dag$ indicates that the result is statistically significant compared to the best baseline SVM+TFIDF.}
\vspace{-2mm}
\label{tab: precision results}
\end{table*}

Since previous investigations on the parameter configuration of word embeddings focus on different tasks such as NER \cite{Godin2015} and dependency parsing \cite{Bansal2014}, their findings may not generalise to Twitter classification tasks. Meanwhile, similar work on Twitter classification tasks \cite{Ebert2015,Severyn2015,Tang2014a} have not studied the impact of background corpora and \textit{Word2Vec} parameters such as context window and dimensionality. Our finding shows that these two factors could affect the classification performance on Twitter classification tasks. Based on experiments on a Twitter election dataset, this paper studies word embeddings when using convolutional neural networks. Using two different types of background corpora, we observe when the type of background corpus aligns with the classification dataset, the CNN classifier can achieve a better performance. In particular, our investigation shows that choosing the correct type of background corpus can potentially cover more vocabulary of the classification dataset. Thus, the alignment between the type of background corpus and classification dataset provides better tweet-level representations. For inferring the best setup of \textit{Word2Vec} parameters (e.g.\ context window and dimensionality), we applied word embeddings with various parameter setup to convolutional neural networks. As a practical guide for a Twitter classification task, word embedding with both large context window and dimension is preferable with a CNN classifier for a better performance.

\section{Acknowledgments}
This paper was supported by a grant from the Economic and Social Research Council, (ES/L016435/1).
\bibliographystyle{abbrv}
\bibliography{sig_Dec2015}

\begin{thebibliography}{10}

\bibitem{bbc2015}
Venezuela opposition politician luis manuel diaz killed.
\newblock http://www.bbc.co.uk/news/world-latin-america-34929332, November
  2015.
\newblock [Accessed: 2016-05-15].

\bibitem{Amati2011}
G.~Amati, G.~Amodeo, M.~Bianchi, G.~Marcone, F.~U. Bordoni, C.~Gaibisso,
  G.~Gambosi, A.~Celi, C.~Di~Nicola, and M.~Flammini.
\newblock {FUB}, {IASI-CNR}, {UNIVAQ} at {TREC} 2011 microblog track.
\newblock In {\em Proc. of TREC}, 2011.

\bibitem{Bansal2014}
M.~Bansal, K.~Gimpel, and K.~Livescu.
\newblock Tailoring continuous word representations for dependency parsing.
\newblock In {\em Proc. of the 52nd ACL conference}, volume~2, pages 809--815,
  2014.

\bibitem{Bengio2003}
Y.~Bengio, R.~Ducharme, P.~Vincent, and C.~Janvin.
\newblock A neural probabilistic language model.
\newblock {\em Journal of machine learning research}, 3:1137--1155, 2003.

\bibitem{Bermingham2011}
A.~Bermingham and A.~F. Smeaton.
\newblock On using {T}witter to monitor political sentiment and predict
  election results.
\newblock In {\em Proc. of SAAIP workshop at IJCNLP}, 2011.

\bibitem{Collobert2008}
R.~Collobert and J.~Weston.
\newblock A unified architecture for natural language processing: Deep neural
  networks with multitask learning.
\newblock In {\em Proc. of the 25th ICML}, pages 160--167, 2008.

\bibitem{Collobert2011}
R.~Collobert, J.~Weston, L.~Bottou, M.~Karlen, K.~Kavukcuoglu, and P.~Kuksa.
\newblock Natural language processing (almost) from scratch.
\newblock {\em Journal of machine learning research}, 12:2493--2537, 2011.

\bibitem{dietterich1998}
T.~G. Dietterich.
\newblock Approximate statistical tests for comparing supervised classification
  learning algorithms.
\newblock {\em Neural computation}, 10(7):1895--1923, 1998.

\bibitem{Ebert2015}
S.~Ebert, N.~T. Vu, and H.~Sch{\"u}tze.
\newblock {CIS}-positive: Combining convolutional neural networks and {SVMs}
  for sentiment analysis in {T}witter.
\newblock In {\em Proc. of the SemEval workshop}, page 527, 2015.

\bibitem{Godin2015}
F.~Godin, B.~Vandersmissen, W.~De~Neve, and R.~Van~de Walle.
\newblock Multimedia {L}ab@ {ACL W-NUT NER} shared task: Named entity
  recognition for {T}witter microposts using distributed word representations.
\newblock In {\em Proc. of the ACL-IJCNLP conference}, page 146, 2015.

\bibitem{Kim2014}
Y.~Kim.
\newblock Convolutional neural networks for sentence classification.
\newblock In {\em Proc. of EMNLP conference}, pages 1746--1751, 2014.

\bibitem{krizhevsky2012}
A.~Krizhevsky, I.~Sutskever, and G.~E. Hinton.
\newblock Imagenet classification with deep convolutional neural networks.
\newblock In {\em Advances in neural information processing systems}, pages
  1097--1105, 2012.

\bibitem{Macdonald2012}
C.~Macdonald, R.~McCreadie, R.~L. Santos, and I.~Ounis.
\newblock From puppy to maturity: Experiences in developing terrier.
\newblock In {\em Proc. of the OSIR workshop at SIGIR}, volume~60, 2012.

\bibitem{Mikolov2013a}
T.~Mikolov, I.~Sutskever, K.~Chen, G.~S. Corrado, and J.~Dean.
\newblock Distributed representations of words and phrases and their
  compositionality.
\newblock In {\em Proc. Advances in neural information processing systems},
  pages 3111--3119, 2013.

\bibitem{Pedregosa2011}
F.~Pedregosa, G.~Varoquaux, A.~Gramfort, V.~Michel, B.~Thirion, O.~Grisel,
  M.~Blondel, P.~Prettenhofer, R.~Weiss, V.~Dubourg, J.~Vanderplas, A.~Passos,
  D.~Cournapeau, M.~Brucher, M.~Perrot, and E.~Duchesnay.
\newblock Scikit-learn: Machine learning in {P}ython.
\newblock {\em Journal of machine learning research}, 12:2825--2830, 2011.

\bibitem{sanderson2010}
M.~Sanderson.
\newblock Test collection based evaluation of information retrieval systems.
\newblock {\em Foundations and Trends in Information Reteieval}, 2010.

\bibitem{Severyn2015}
A.~Severyn and A.~Moschitti.
\newblock {UNITN}: Training deep convolutional neural network for {T}witter
  sentiment classification.
\newblock In {\em Proc. of the 9th SemEval workshop}, pages 464--469, 2015.

\bibitem{severyn2015d}
A.~Severyn, M.~Nicosia, G.~Barlacchi, and A.~Moschitti.
\newblock Distributional neural networks for automatic resolution of crossword
  puzzles.
\newblock In {\em Proc. of {ACL-IJCNLP} conference}, 2015.

\bibitem{Tang2014a}
D.~Tang, F.~Wei, N.~Yang, M.~Zhou, T.~Liu, and B.~Qin.
\newblock Learning sentiment-specific word embedding for {T}witter sentiment
  classification.
\newblock In {\em Proc. of the 52nd ACL conference}, volume~1, pages
  1555--1565, 2014.

\bibitem{Wang2015}
P.~Wang, J.~Xu, B.~Xu, C.-L. Liu, H.~Zhang, F.~Wang, and H.~Hao.
\newblock Semantic clustering and convolutional neural network for short text
  categorization.
\newblock In {\em Proc. of the 53rd {ACL-IJCNLP} conference}, volume~2, pages
  352--357, 2015.

\end{thebibliography}
%
%

\end{document}